\documentclass[letterpaper, paper,11pt]{IWSCFF}		

\usepackage{bm}
\usepackage{amsmath}
\usepackage[colorlinks=true, pdfstartview=FitV, linkcolor=black, citecolor= black, urlcolor= black]{hyperref}
\usepackage{footnpag}			      	
\usepackage[sort,comma,numbers,super]{natbib}
\usepackage{subfig}
\usepackage{algorithm}
\usepackage{algpseudocode}
\usepackage{placeins}

\PaperNumber{17-33}

\begin{document}

\title{Differential drag control scheme \protect\\
	for large constellation of Planet satellites \protect\\
and on-orbit results}

\author{Cyrus Foster\thanks{Orbit Mechanic, AIAA member, cyrus.foster@planet.com}, 
James Mason\thanks{VP Missions, AIAA member, james.mason@planet.com},
Vivek Vittaldev\thanks{Missions Operations Analyst, AIAA member, vivek@planet.com},
Lawrence Leung\thanks{ADCS Engineer, AIAA member, lawrence@planet.com}, \\
Vincent Beukelaers\thanks{ADCS Engineer, AIAA member, vincent@planet.com},
Leon Stepan\thanks{Missions Operations Analyst, AIAA member, leon.stepan@planet.com} and
Rob Zimmerman\thanks{Systems Engineer, AIAA member, rob@planet.com}}

\maketitle{} 		

\begin{abstract}
A methodology is presented for the differential drag control of a large fleet of 
propulsion-less satellites deployed in the same orbit. The controller places satellites into a constellation with specified angular offsets and zero-relative speed. 
Time optimal phasing is achieved by first determining an appropriate relative placement, i.e. the order of the satellites. A second optimization problem is then solved as a large coupled system to find the drag command profile required for each satellite. 
The control authority  is the available ratio of low-drag to high-drag ballistic coefficients of the satellites when operating in their background mode. The controller is able to successfully phase constellations with up to 100 satellites in simulations. On-orbit performance of the controller is demonstrated by phasing the Planet Flock 2p constellation of twelve cubesats launched in June 2016 into a 510 km sun-synchronous orbit.
\end{abstract}

\section{Introduction}

\begin{figure}[!ht]
	\centering\includegraphics[width=2.0in]{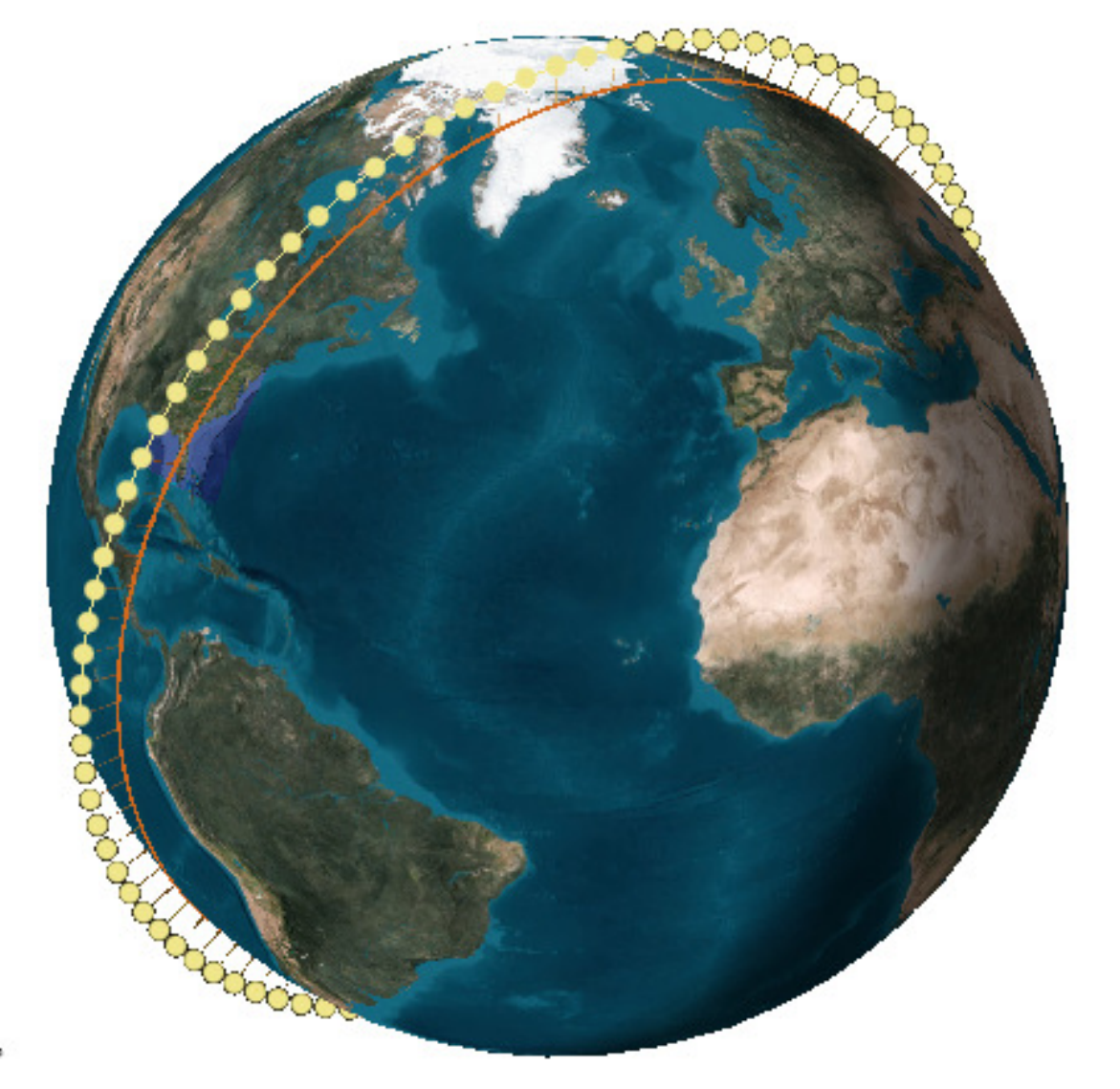}
	\caption[caption]{A line scanner configuration of satellites uniformly spread over the orbit}
	\label{fig:linescannerfull}
\end{figure}

Due to easier access to space in recent years, there have been an increasing number of satellites launched. In addition to more launches, multiple satellites are deployed into orbit in one launch. The resent launch by the Indian Space Research Organisation (ISRO), for example, carried 104 satellites into orbit in February 2017. Other launches with a large number of satellites were a Russian Dnepr launch with 37 satellites, and an Orbital Antares rocket with 34 satellites onboard in 2004. Such launches are optimal for large constellations of small satellites that are in the same orbit but spread out in mean anomaly. The customized or randomized deployment velocities provided by the launcher can help in the spreading of the satellites. It is, unfortunately, not possible to first spread the satellites and then maintain relative phasing by only using the deployment velocities. To realize a fully spread and operationally useful constellation as shown in Figure~\ref{fig:linescannerfull}, some propulsive capability is required.    

A constellation launched together typically consists of small homogeneous satellites. Adding propulsion to these satellites is an undertaking in terms of engineering, cost, and regulations. It might not  be possible to fit the required propulsion subsystem for the desired $\Delta v$ due to size and weight constrains. Atmospheric drag, however, can be used to provide the relative velocities required to phase the constellation without additional mass if the altitude is sufficiently low.
Differential drag changes the drag characteristics of space objects in order to control relative accelerations. The effective surface area perpendicular to the velocity vector acts as the actuator by changing the Ballistic Coefficient (BC).

Differential drag was first proposed for station keeping of a pair of satellites
in 1989 \citep{leonard} by using drag plates. The control freedom is the angle of attack of the drag plates at $0^{\circ}$ or $90^{\circ}$ and the control law is based on the
linearized Clohessy-Wiltshire equations of motion~\citep{ClohessyWiltshire}. A similar strategy using a drag plate and including $J_2$ perturbation for satellite rendezvous is demonstrated by Bevilacqua~\citep{Bevilacqua2008} using the Schweighart-Sedwick linearized equations of motion~\citep{Schweighart-Sedwick}.  Rendezvous using a combination of differential drag for the first phase and low-thrust propulsion for the second precise phase is shown~\citep{Bevilacqua2009}. Differential drag has also been combined with other perturbations such as Solar Radiation Pressure (SRP) for formation maneuvers~\citep{DiffDragSRP}. 

Various types of controllers have been used to create trajectories such as an optimal control approach~\citep{DellElceOptimal}, Lyapunov based controllers~\citep{BevilacquaLyapunovJGCD}, and sliding mode control~\citep{VarmaDiffDragSlidingMode}. The extension of the two satellite problem into a multiple satellite problem has been carried out previously~\citep{MattHarrisOptimalDrag, kumar, horsley}.

Differential drag has also been used for collision avoidance~\citep{MishneDiffDragCollisionAvoidance, HuangDiffDragCollision}. For satellites using differential drag, uncertainties in the atmosphere models have a noticeable effect on the uncertainty propagation of the satellites~\citep{MazalDiffDragUncertainty, BenYaacovDiffDragUncertainty}. Another possible application of differential drag is to minimize the post-deployment probability of collision of nanosatellties~\citep{AtchisonDiffDragDeployment}.

To date, the Aerospace Corporation has demonstrated the use of differential drag on-orbit with small
satellites, most recently with the AeroCube-4 mission in 2013 \citep{aerocube}.
This three-satellite mission was deployed into a 480 x 780 km orbit and operators were able
to control along-track spacing of the satellites by selectively retracting solar panel
wings.
Orbcomm, with its constellation of 35 satellites in a 720 km altitude circular orbit,
has also used differential drag to save propellant on its thrust-based station-keeping system \citep{orbcomm}.

The work presented in this paper shows an end to end implementation of a differential drag controller that phases out and performs station keeping on a constellation of satellites. The proposed controller converts the phasing problem into two separate optimization problems. The phase called \emph{Reslotting} is only relevant when the constellation has more than two spacecraft where the relative phasing of the satellites is determined. The atmospheric density is allowed to be time varying, as is the case for real missions. Almost all prior work in differential drag, except for Guglielmo et al.~\citep{GuglielmoDiffDragDensityForecasting}, assumes a constant density. Intra-constellation coupling is also included when computing the control. Coupling is considered by Harris and A\c{c}ikme\c{s}e~\citep{MattHarrisOptimalDrag} for up to five spacecraft with constant atmospheric drag. It is, however, unclear how this method scales up to 100 satellites. Li and Mason~\citep{li} treated a degenerate case of the multi-spacecraft problem limited to five satellites where the reference satellite performs no high drag maneuver. In addition to simulation results, on-orbit results of the successful phasing of 12 cubesats and the ongoing phasing of 88 cubesats is shown.

\section{Differential-Drag Controller}

\label{section:controller}

\begin{figure}[!ht]
\centering\includegraphics[height=2.0in]{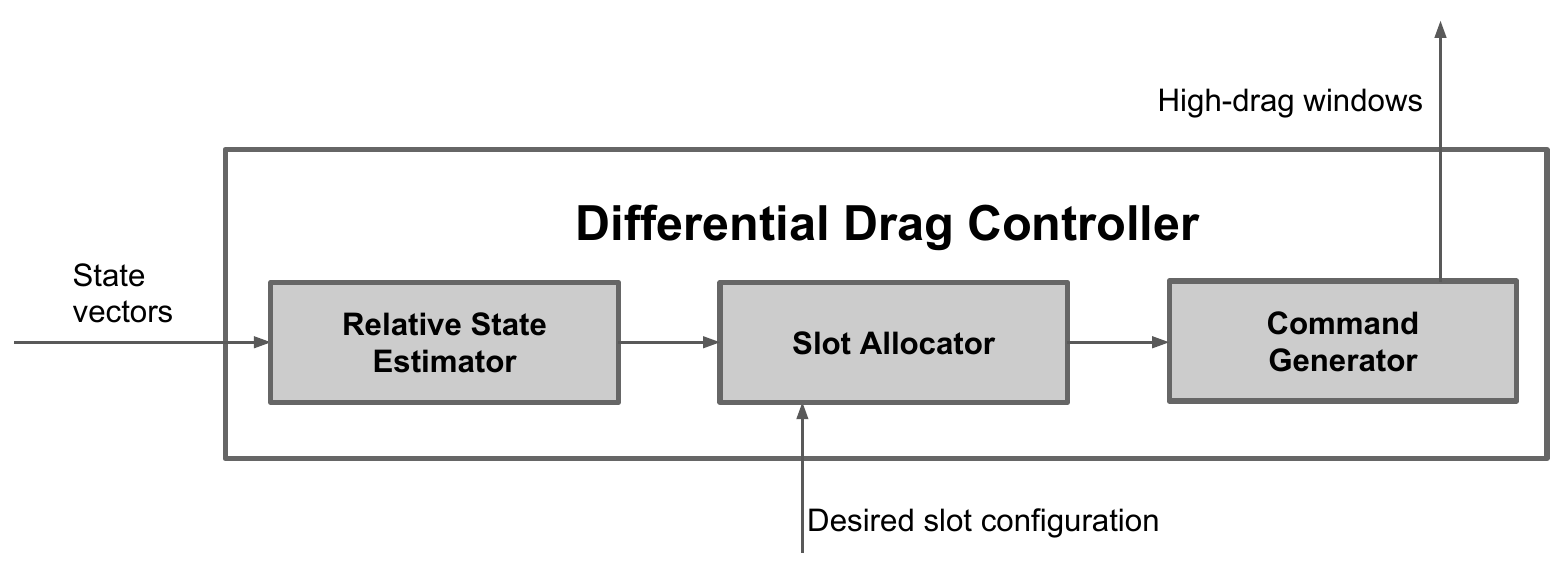}
\caption[caption]{Controller produces high-drag windows from state vectors}
\label{fig:controller}
\end{figure}

As illustrated in Figure \ref{fig:controller}, the controller consists of three components.
It first uses Cartesian position-velocity state vectors generated by orbit determination 
to estimate the relative motion of the satellites in the along-track direction.
Two optimization problems aiming to minimize the total phasing time are then solved sequentially: 
\begin{enumerate}
	\item Slot Allocator: allocating each satellite to specific slots
	\item Command Generator: finding appropriate high-drag windows required to reach the slots
\end{enumerate}

\subsection{Relative State Estimator}

\begin{figure}[!ht]
\centering\includegraphics[height=2.5in]{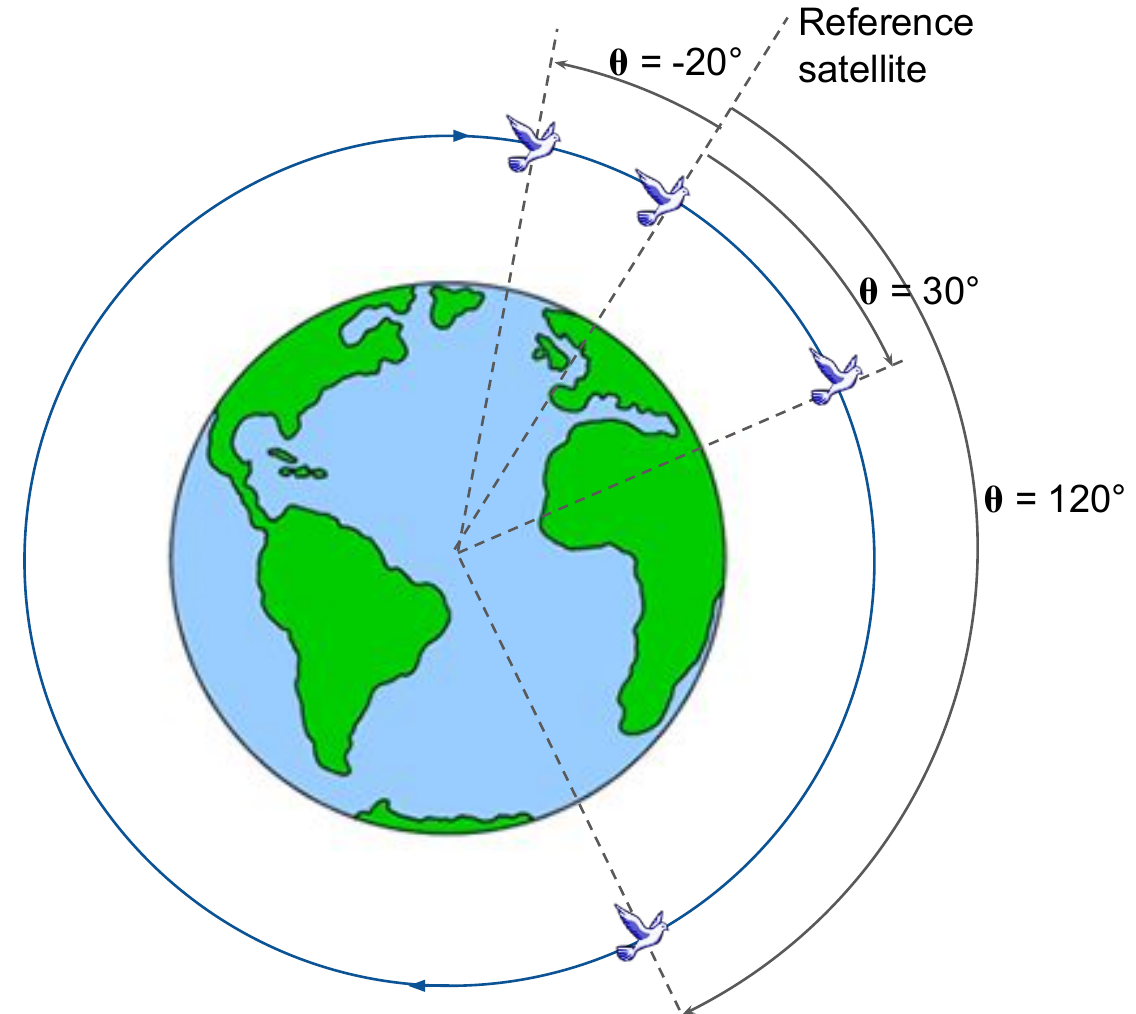}
\caption[caption]{Relative state $\theta$ reduces complexity of control problem to along-track space}
\label{fig:relative_angle}
\end{figure}

Since differential drag only controls the along-track dynamics of the satellite,
the independent set of cartesian state vectors are reduced to relative 
Earth-centered angles shown in Figure \ref{fig:relative_angle} and their time derivatives.

\subsubsection{Initial Condition}
Relative angles $\theta$ are calculated between each satellite and a reference satellite.
This reference can be chosen to be any of the satellites but by convention 
the one that starts in the lowest orbit is used.

$\theta_{osc}$ is defined 
as the osculating Earth-centered relative angle between each satellite $i$ and
the reference's position vectors $R$, defined over $[0..2\pi)$ in the along-track direction 
(Equation \ref{eq:theta_osc}).
The half-plane ambiguity is solved with the reference pole vector
$\vec{N} = || \vec{R}_{ref} \times \vec{V}_{ref} ||$.
It is noted that the reference satellite's $\theta$ and $\dot{\theta}$ are always 0 by
definition.

\begin{equation}
\label{eq:theta_osc}
\theta_{osc,i} = angle_{[0..2\pi)}(\vec{R}_{ref}, \vec{R}_i)
\end{equation}

\begin{figure}%
\centering
\subfloat[Relative state $\theta_0$ and $\dot{\theta_0}$ from osculating motion]{{
\includegraphics[width=.46\linewidth]{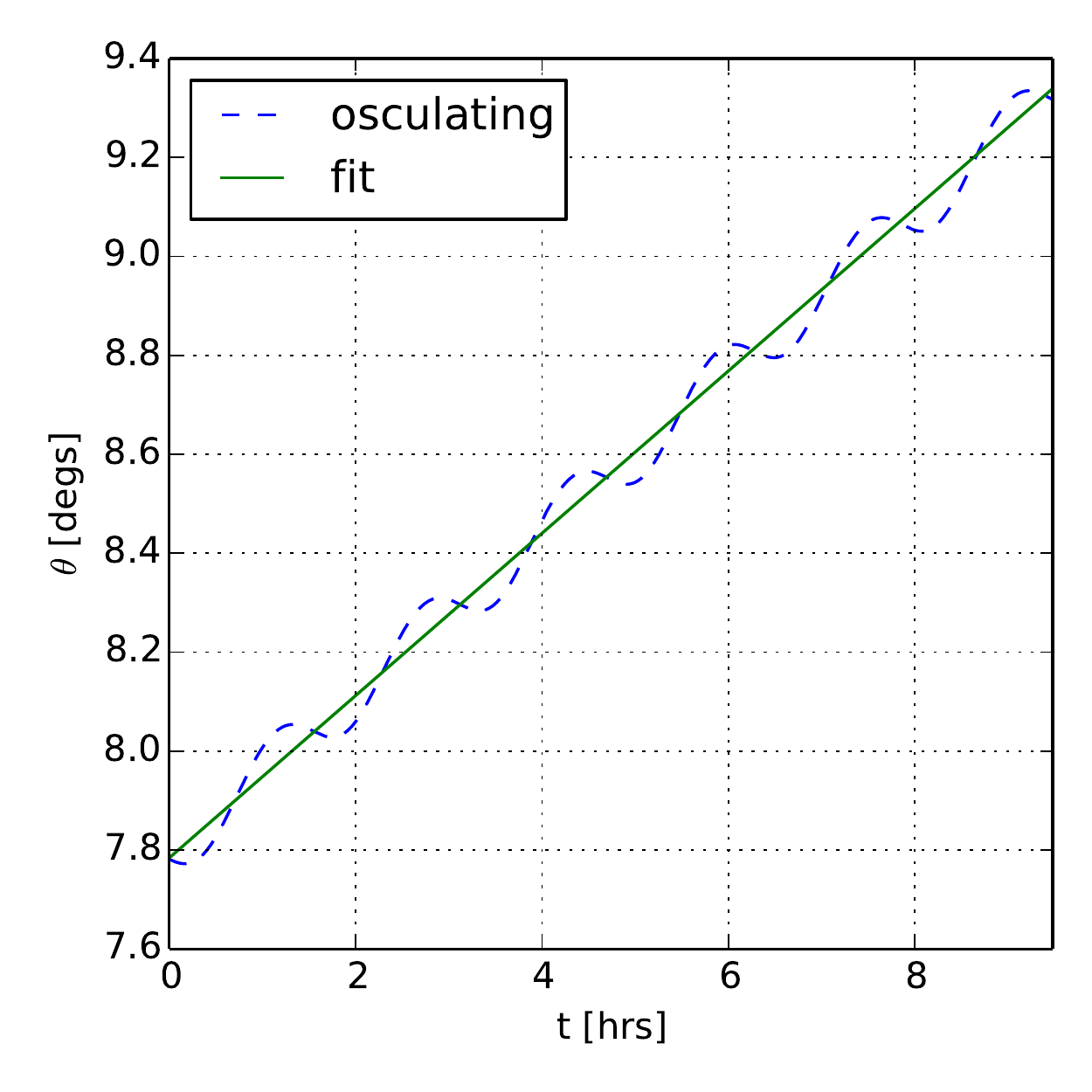} }}%
\qquad
\subfloat[Mean atmosphere density encountered by satellite,
and hence control authority for differential drag,
can fluctuate with time]{{
\includegraphics[width=.46\linewidth]{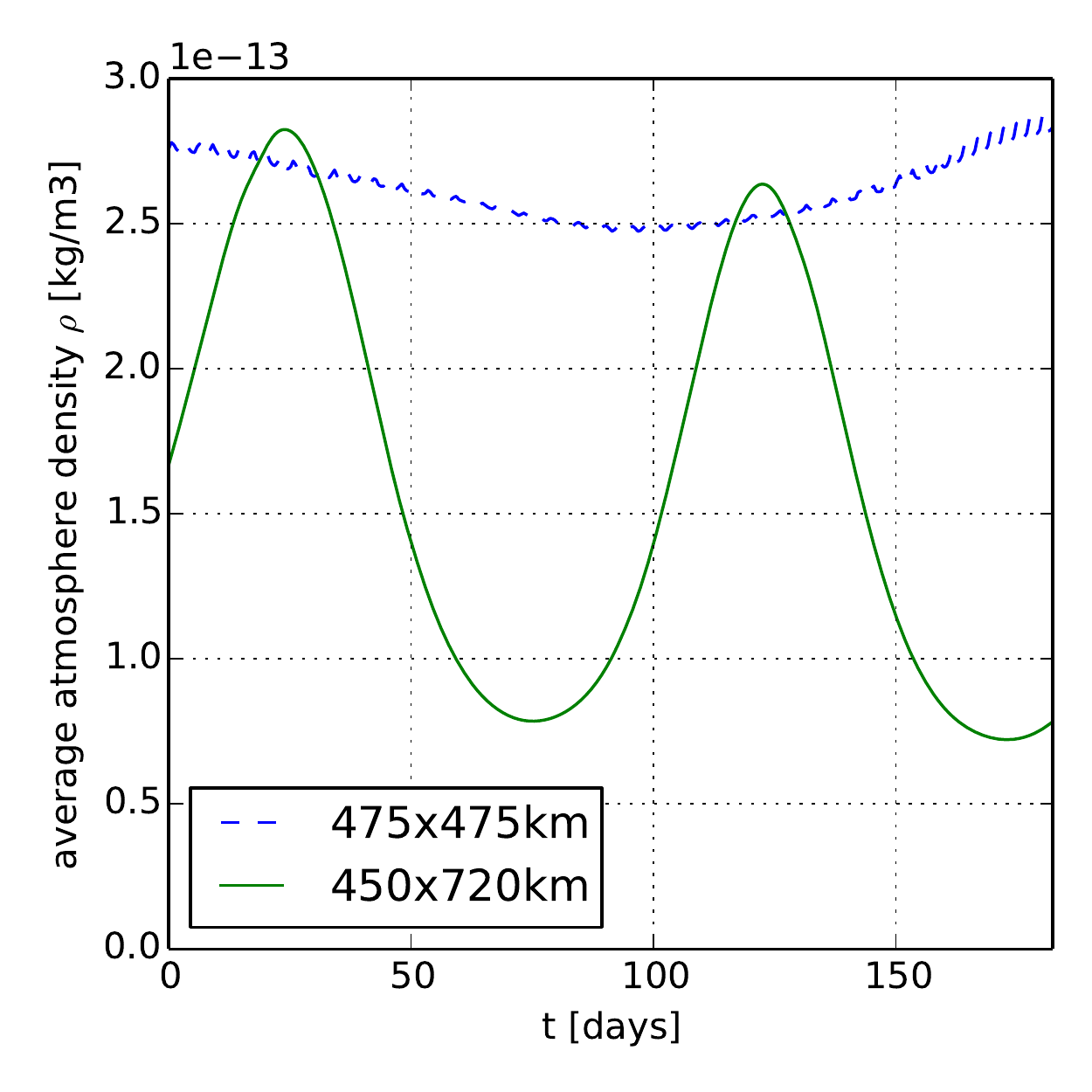} }}%
\caption{Relative motion is estimated from propagated satellite state vectors}%
\label{fig:rel_motion}%
\end{figure}

The initial state consisting of $\theta$ and its time derivative, 
the angular velocity $\dot{\theta}$, are calculated
as mean elements by fitting a line through the time history of
$\theta_{osc}$ via least-squares as shown in Figure \ref{fig:rel_motion}a.
A 1-day fit period proves to be sufficient for capturing the motion of a typical LEO SSO orbit (400-600 km).
To produce this time history, orbits are propagated under a 10x10 gravity and
NRL-MSIS00 atmosphere force model.

\subsubsection{Control authority}
To model the dynamics of satellites in high-drag, $\ddot{\theta}$ is also calculated,
the relative angular acceleration of one satellite in high-drag mode relative to
the reference satellite in low-drag.
$\ddot{\theta}$ is a measure of available control authority, 
and is calculated across the simulation time horizon to effectively model
the time-varying nature of control authority due to osculating motion of perigee and apogee
and forecasted solar flux variations.
This time-varying nature is important to capture when planning months ahead,
and especially so for elliptical orbits where perigee and apogee vary
cyclically over month timescales due the Earth's gravity field 
(Figure \ref{fig:rel_motion}b).

The mean dynamic pressure encountered by the reference satellite $q_{\text{ref}}$ throughout a discretized time period $\Delta t$ at time $k$  is first obtained using the atmosphere model with the latest space weather data.
\begin{equation}
q_{\text{ref},k} = \frac{1}{2} \overline{ \rho v^2 }
\end{equation}

Next, the $\Delta V$ during the discretized time period is calculated using 
the observed ballistic coefficients $BC = \frac{m}{C_d A}$ in low and high-drag modes,
$BC_L$ and $BC_H$ respectively:

\begin{equation}
\Delta V = q_{\text{ref},k}\Big(\frac{1}{BC_L} - \frac{1}{BC_H}\Big) \Delta t
\end{equation}

The mean relative velocity $\dot{y}$ between the satellites is then evaluated,
using the secular along-track component 
of the solution to Hill's equations for relative motion \citep{vallado}:

\begin{equation}
\dot{y} \approx - 3 \dot{y}_0 = - 3 \Delta V = 
3 q_{\text{ref},k}\Big(\frac{1}{BC_H} - \frac{1}{BC_L}\Big) \Delta t
\end{equation}

Finally, the angular acceleration $\ddot{\theta}_k$ is evaluated during this
discretized period,
converting to angular coordinates using the reference
satellite's mean semi-major axis at epoch $\overline{a}_{\text{ref}}$:

\begin{equation}
\ddot{\theta}_k = \frac{\dot{y}\Delta t}{\overline{a}_{\text{ref}}} =
\frac{3 q_{\text{ref},k}}{\overline{a}_{\text{ref}}}\Big(\frac{1}{BC_H} - \frac{1}{BC_L}\Big)
\end{equation}

\subsubsection{Dynamics}
The along-track dynamics of a constellation of $n$ satellites subject to time-discretized
binary control inputs (low-drag or high-drag) can now be summarized by
$n-1$ systems of Equation \ref{eq:dynamics} for each satellite-reference pair.
Note these systems are coupled since they all share the reference's
commands $u_{ref}$.

\begin{subequations}
\label{eq:dynamics}

\begin{equation}
\begin{bmatrix} \theta \\ \dot{\theta} \end{bmatrix}_{k+1} = 
\begin{bmatrix} 1 & \Delta t \\ 0 & 1 \end{bmatrix}
\begin{bmatrix} \theta \\ \dot{\theta} \end{bmatrix}_k + 
\begin{bmatrix} 0 \\ B_k(u_k) \end{bmatrix}
\end{equation}

where the control inputs $B_k$ are evaluated over the discretization period $\Delta t$
with the combined angular acceleration from the two satellites:
\begin{equation}
B_k(u_k) = 
(-\ddot{\theta}_{ref,k} u_{ref,k} + \ddot{\theta}_{sat,k} u_{sat,k})\Delta t
\end{equation}

and high-drag commands $u_{i,k}$ are defined by:
\begin{equation}
u_{i,k} =
\begin{cases} 
0 \quad \text{if \textit{i} is in low-drag at} \quad t_k \\ 
1 \quad \text{if \textit{i} is in high-drag at} \quad t_k
\end{cases}
\end{equation}

\end{subequations}

\subsection{Slot Allocator}
\begin{figure}%
\centering
\subfloat[random slot order: $\Delta t_{\text{phase}} = 90$ days]
{{\includegraphics[width=.40\linewidth]{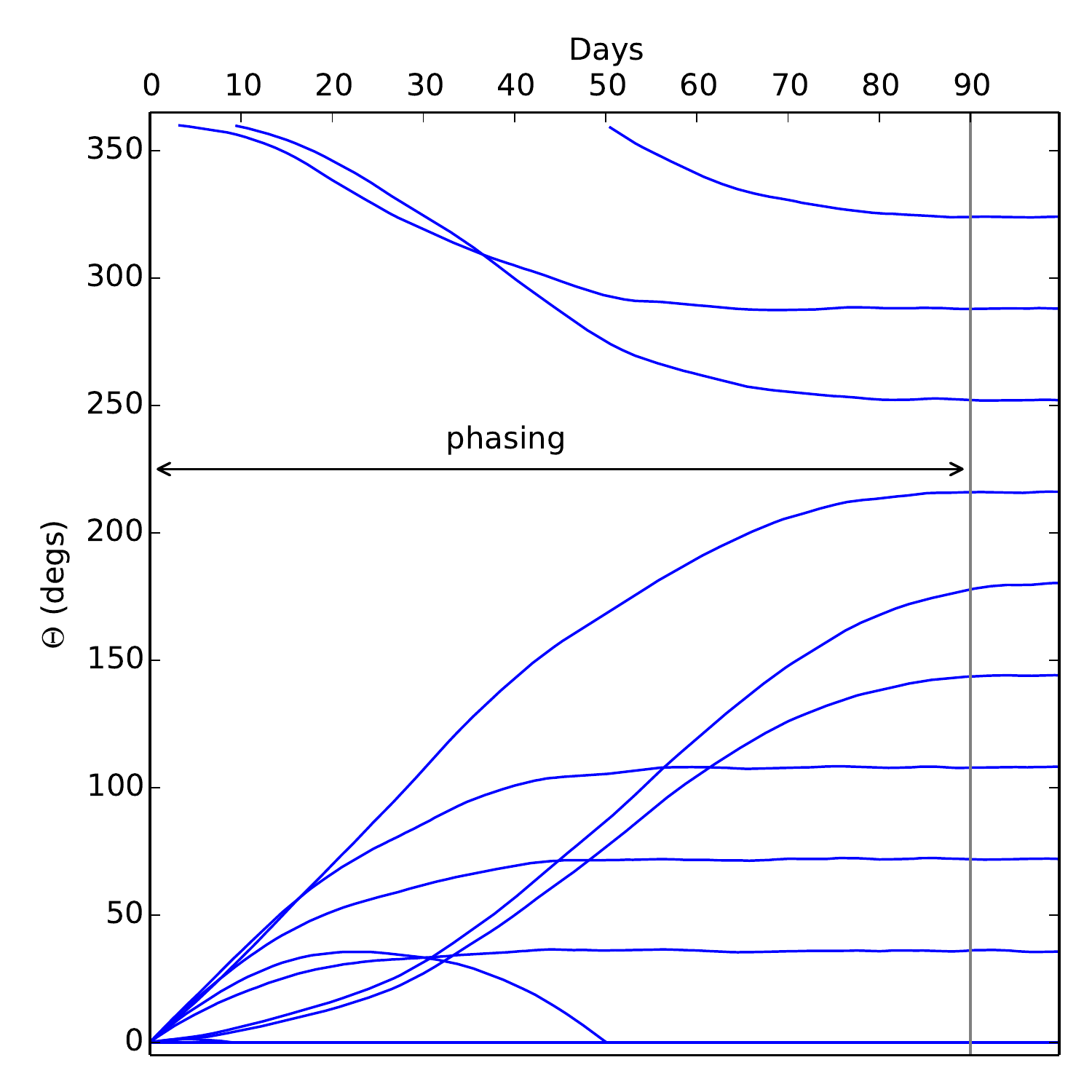} }}%
\qquad
\subfloat[optimal slot order: $\Delta t_{\text{phase}} = 70$ days]
{{\includegraphics[width=.40\linewidth]{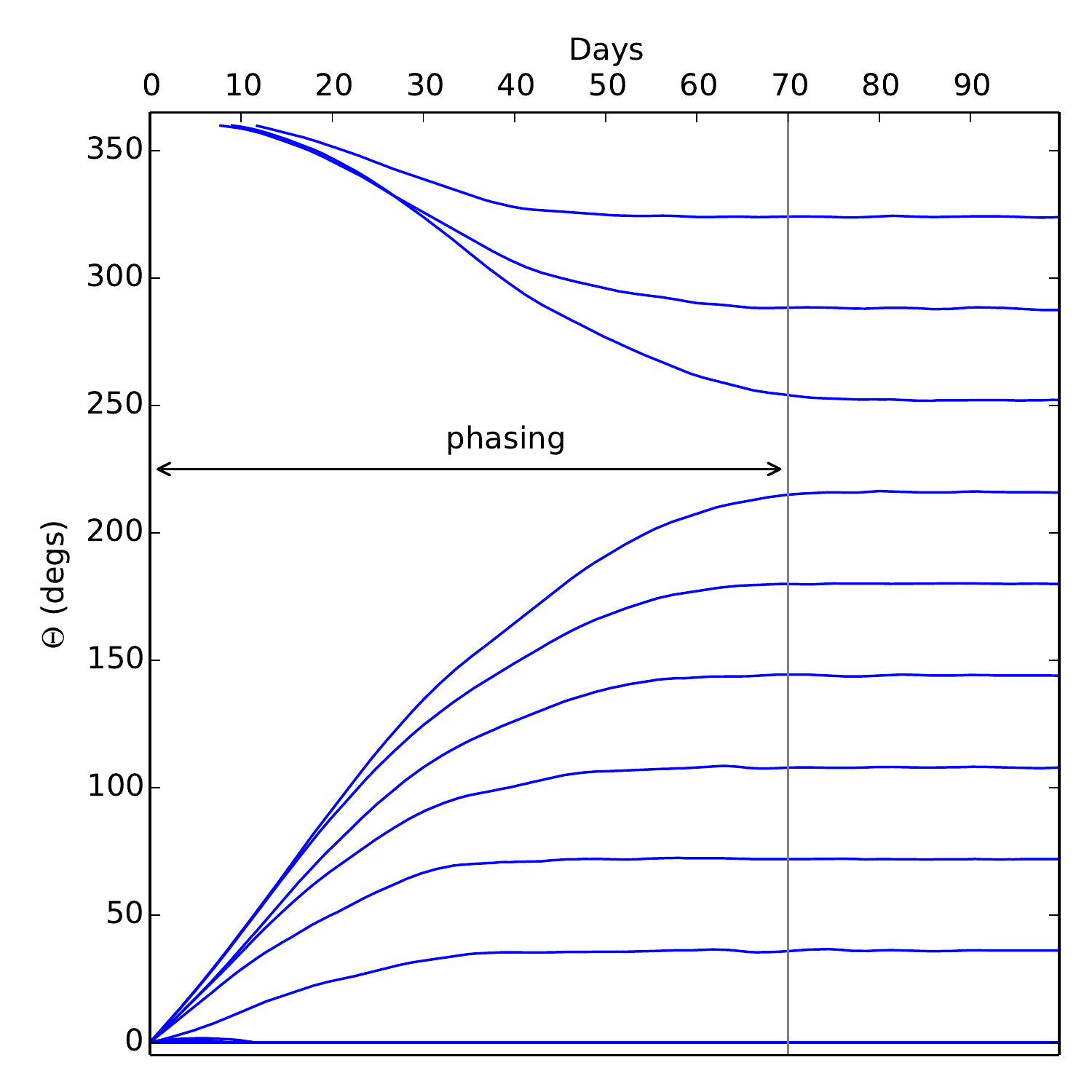} }}%
\caption{Allocating slots judiciously significantly lowers
the time required to phase a constellation.}%
\label{fig:slot_noslot}%
\end{figure}

With the control problem reduced to optimization in relative-motion $\theta$ space,
satellite ordering relative to one another is first determined to
minimize predicted phasing time. 
The Dove satellites are interchangeable, 
so optimization is targeted to seek the lowest phasing time that achieves 
the desired slot vector irrespective of satellite order. 
As shown in Figure \ref{fig:slot_noslot},
the judicious assignment of satellites to slots
is important for minimizing the time to phase of the constellation.
Figure \ref{fig:slot_noslot}a shows a phasing attempt when target slots
are assigned randomly,
while Figure \ref{fig:slot_noslot}b starts with the same initial condition
but slots have been allocated optimally to minimize time to phase.

\subsubsection{Slots}
\begin{figure}[!ht]
\centering\includegraphics[height=2.0in]{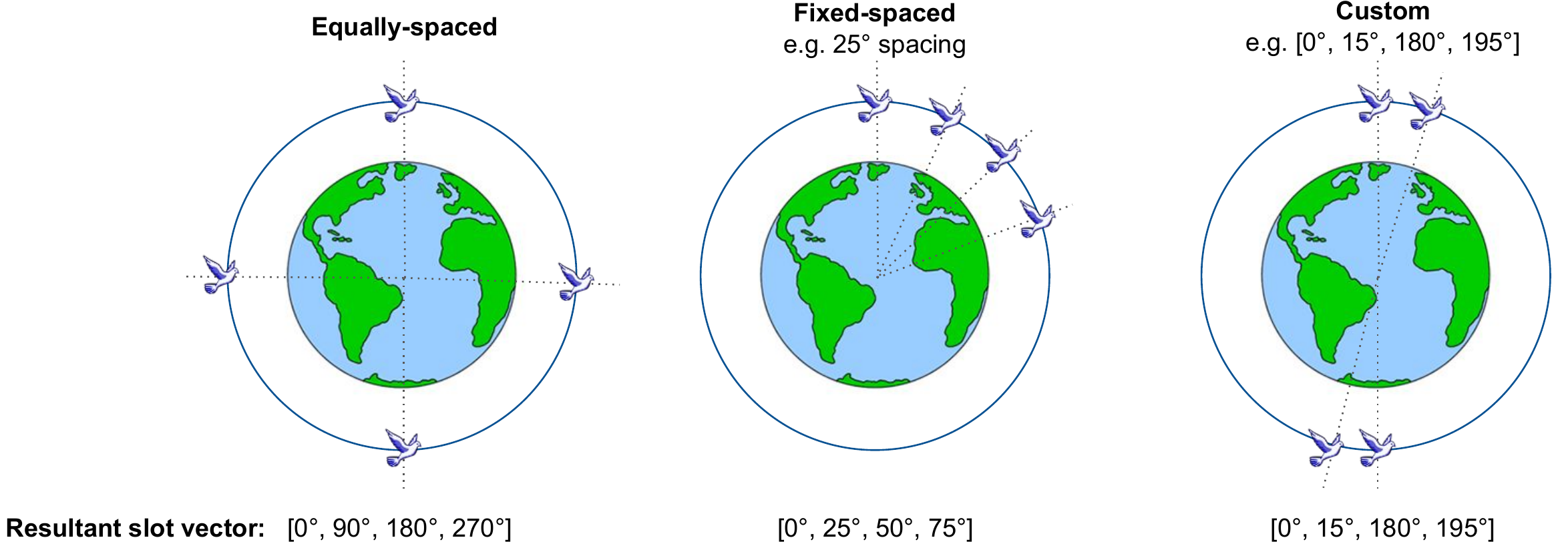}
\caption[caption]{Slot vector describes desired relative placement of satellites}
\label{fig:slots}
\end{figure}

A vector of slots is defined, where a slot is a desired $\theta_f$ for the
final state of slot with $\dot{\theta}_f = 0$.
Various examples of slot configurations are shown in Figure \ref{fig:slots}.
Equally-spaced satellites make sense for minimizing imaging swath overlap and ground
station conflicts, while the fixed-spaced configuration is ideal for realizing a
line scanner with adjacent swaths.

\subsubsection{Flip-flop solution}

Since the objective is to minimize phasing time, there is a need to estimate the
amount of time required to achieve a certain slot configuration 
$\begin{bmatrix} \theta & \dot{\theta} \end{bmatrix}^T_f$
given the initial condition 
$\begin{bmatrix} \theta & \dot{\theta} \end{bmatrix}^T_0$
of each non-reference satellite.
A method is introduced for solving the phasing time of a two-satellite 
constellation open-loop; 
to be used subsequently in the slot allocation optimizer.

\begin{figure}[!ht]
\centering\includegraphics[width=2.4in]{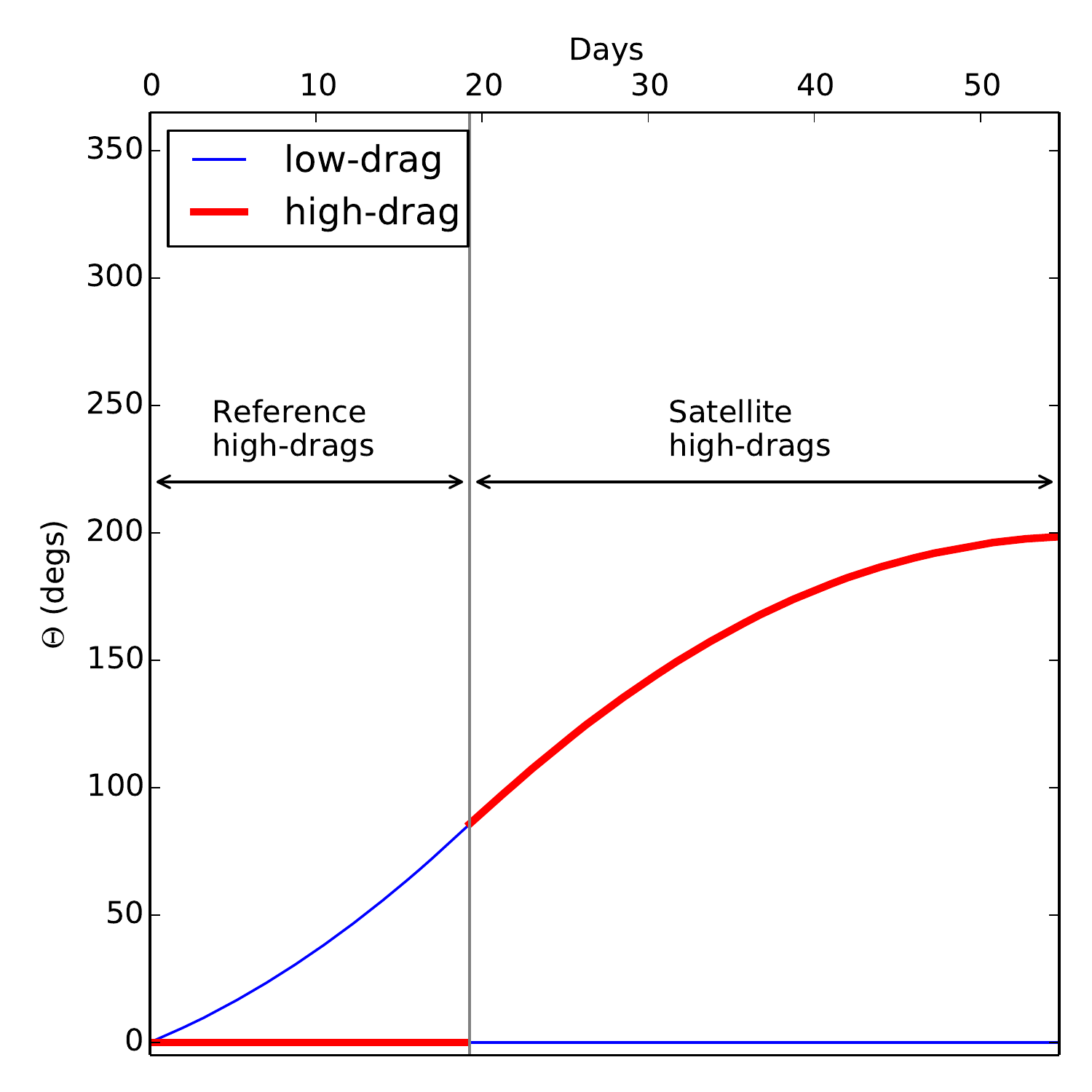}
\caption[caption]{Flip-flop solution provides the time-optimal schedule for phasing two satellites}
\label{fig:flipflop}
\end{figure}

The time-optimal solution for phasing two-satellites always has two phases 
(Figure \ref{fig:flipflop}):
one satellite high-drags for a duration $\Delta t_A$ while the other low-drags (flip),
then the roles reverse for another duration $\Delta t_B$ (flop).
The solution to this control problem (duration of each phase, which satellite starts first)
can be found analytically if assuming time-invariant $\ddot{\theta}$ (see Appendix B). 
The general problem can also be solved numerically via a line search as outlined in
Algorithm \ref{alg:flipflop},
where the relative motion is propagated with the discretized equations of motion from Equation \ref{eq:dynamics}
to account for time-varying $\ddot{\theta}$.

\begin{algorithm}[H]
\caption{Flip-flop solution}
\label{alg:flipflop}
\begin{algorithmic}[1]
\Procedure{Solve two-satellite flip-flop problem}{}
\State try both order combinations for first sat to high-drag 
(one has a bracketed root in line search)
\While{line searching to find $\Delta t_A$ such that 
$\theta_f = \theta_{f,desired}$}
\State prop via Eq \ref{eq:dynamics} 
first sat in high-drag for $\Delta t_A$ then second in high-drag until 
$\dot{\theta}_f = \dot{\theta}_{f,desired}$
\EndWhile
\State output $\Delta t_{\text{phase}} = \Delta t_A + \Delta t_B$, 
and order in which
satellites high-drag.
\EndProcedure
\end{algorithmic}
\end{algorithm}

\subsubsection{Optimizer}
The optimal slot configuration is solved for by selecting the order which minimizes
the largest $\Delta t_{\text{phase}}$ of each reference-satellite pair.
Algorithm \ref{alg:slot} describes an implementation using simulated annealing
\citep{simulated_annealing}
wherein random perturbations are made to the slot configuration and kept if
it satisfies an acceptance probability described in equation \ref{eq:sa}.
It is found that values of $k_{max} = 10^6$ and $t_0 = 100$ produce good results
for typical differential drag scenarios.

\begin{algorithm}[H]
\caption{Slot Allocator}
\label{alg:slot}
\begin{algorithmic}[1]
\Procedure{Obtain slot configuration for minimum time phasing}{}
\For{$k$ through $k_{max}$}
\State create $\text{slot}_{new}$ from $\text{slot}$ by randomly swapping two satellites
\State compute $\Delta t$ to phase for all reference-sat pairs with Algorithm \ref{alg:flipflop}
\If{$P(\Delta t_{\text{max,old}}, \Delta t_{\text{max,new}}, k, k_{max}) \geq
\text{random}(0, 1)$}
\State slot $\gets \text{slot}_{new}$
\EndIf
\EndFor
\EndProcedure
\end{algorithmic}
\end{algorithm}

\begin{equation}
\label{eq:sa}
P(\text{cost}, \text{cost}_\text{new}, k, k_{max}) = 
\begin{cases} 
1 \quad 
\text{if cost}_\text{new} < \text{cost}  
\\ 
\exp{(\frac{\text{cost}-\text{cost}_\text{new}}{t})} \quad 
\text{if cost}_\text{new} \geq \text{cost with} \quad
t = t_0 ( 1 - \frac{k}{k_{max}} )
\end{cases}
\end{equation}

If the condition
$\Delta t_{\text{max,old}} = \Delta t_{\text{max,new}}$
is reached, 
because the reference-satellite pair with the maximum phasing time is 
the same in the nominal and perturbed slot configurations, 
one then backs up to comparing the second greatest
$\Delta t_{\text{phase}}$ (or next greatest if still equal).

When computing $\Delta t$ to phase to a particular slot, 
one also solves for which $2\pi$ offset to use by trying several 
neighboring $2\pi$ multiples and selecting the one with lowest phasing time.

\subsection{Command Generator}

\begin{figure}%
\centering
\subfloat[without commands]{{\includegraphics[width=.40\linewidth]{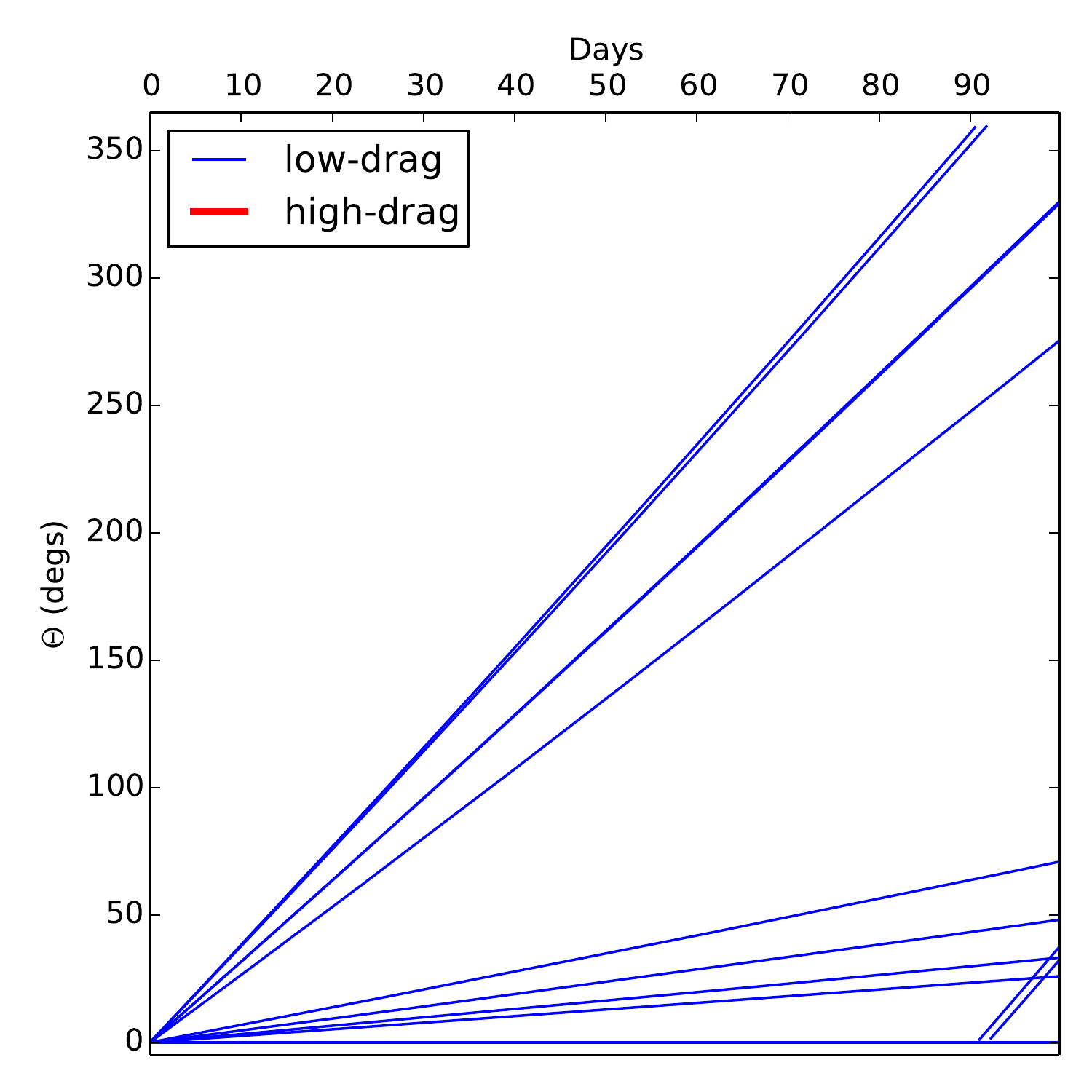} }}%
\qquad
\subfloat[with commands]{{\includegraphics[width=.40\linewidth]{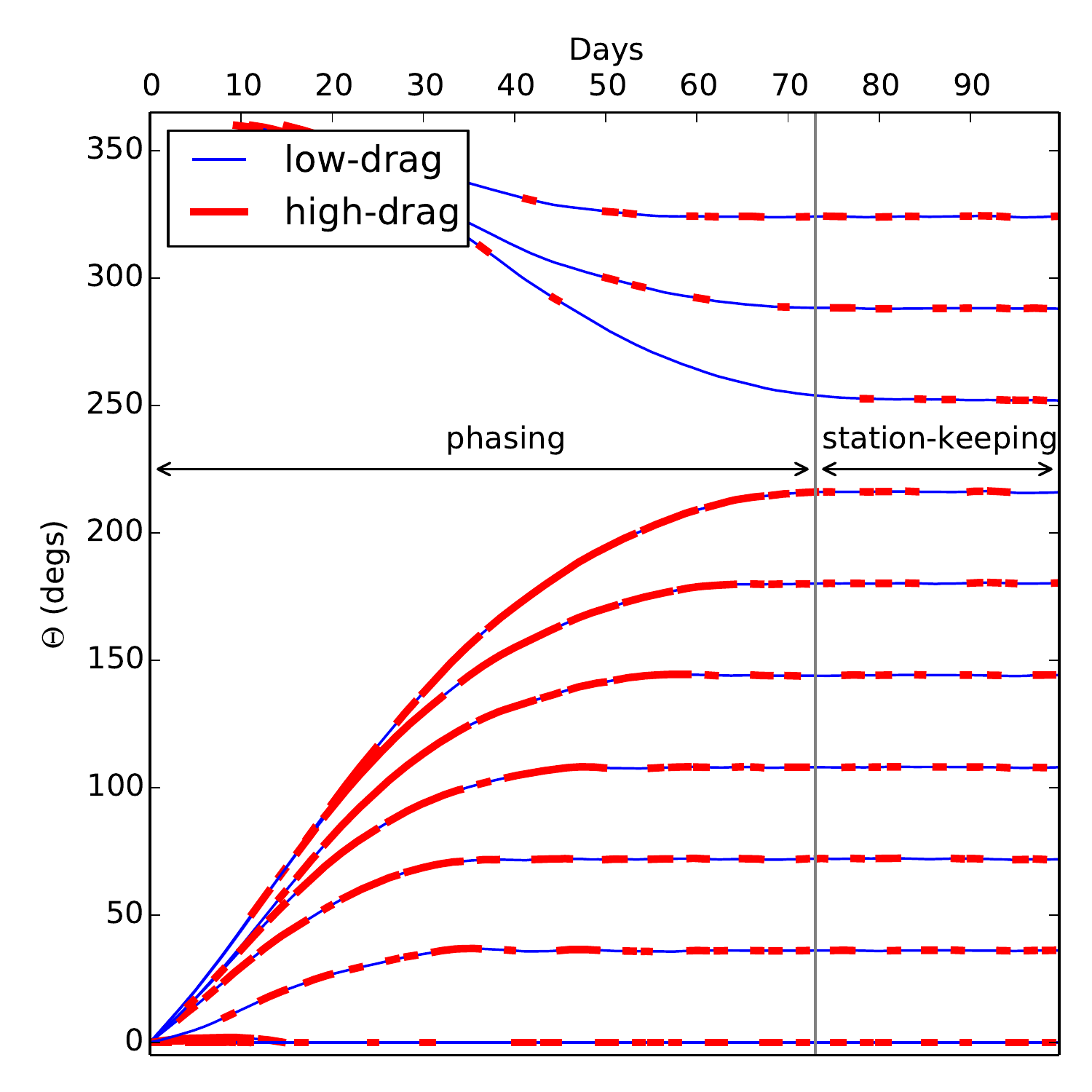} }}%
\caption{Time-discretized high-drag commands are assigned to achieve
desired slots}%
\label{fig:commands}%
\end{figure}

Now that a final desired state $\theta_f$ is determined for each satellite,
a set of high-drag commands is needed to guide the satellites
to their desired slots in minimum time (Figure \ref{fig:commands}).

High-drag commands are discretized for simulation with Equation \ref{eq:dynamics}
across a time horizon that sufficiently captures the required settling time
as estimated by the flip-flop solutions.
The initial guess for the command matrix $u$ comes from the superposition
of commands from flip-flop solutions to each reference-satellite sub-problem,
or alternatively from the commands of the last time the controller was run.
$u$ is then perturbed to converge satellites to their assigned slots via
simulated annealing per the procedure described in Algorithm \ref{alg:command_generator}.

\begin{algorithm}[H]
\caption{Command generator}
\label{alg:command_generator}
\begin{algorithmic}[1]
\Procedure{Obtain high-drag commands to target slots in minimum time}{}
\For{$k$ through $k_{max}$}
\State create $u_{new}$ with randomly flipped command $u_{sat=i,t=k}$
\State compute finals states $\theta_{i,f}$ under $u_{new}$ with eq. \ref{eq:dynamics}
\State compute $\text{cost}_{new} = \sum_{i}(\theta_{i,f} - \theta_{i,desired})^2$
\If{$P(\text{cost}_{old}, \text{cost}_{new}, k, k_{max}) \geq
\text{random}(0, 1)$}
\State $u \gets u_{new}$
\EndIf
\EndFor
\EndProcedure
\end{algorithmic}
\end{algorithm}

During the station-keeping phase, the default high-drag mode is replaced with one
that performs high-drag only a fraction of the time (e.g. 25\%)
to minimize the lifetime lost to drag since full control authority is no longer required.
This also has the desireable side effect of shortening the limit cycle (station-keeping phasing error)
when the time discretization period is kept constant.

\section{Results}
\label{section:results}
\begin{figure}
	\centering
	\subfloat[Orthographic projections with cross-sectional areas.]
	{{\includegraphics[height=2.5in]{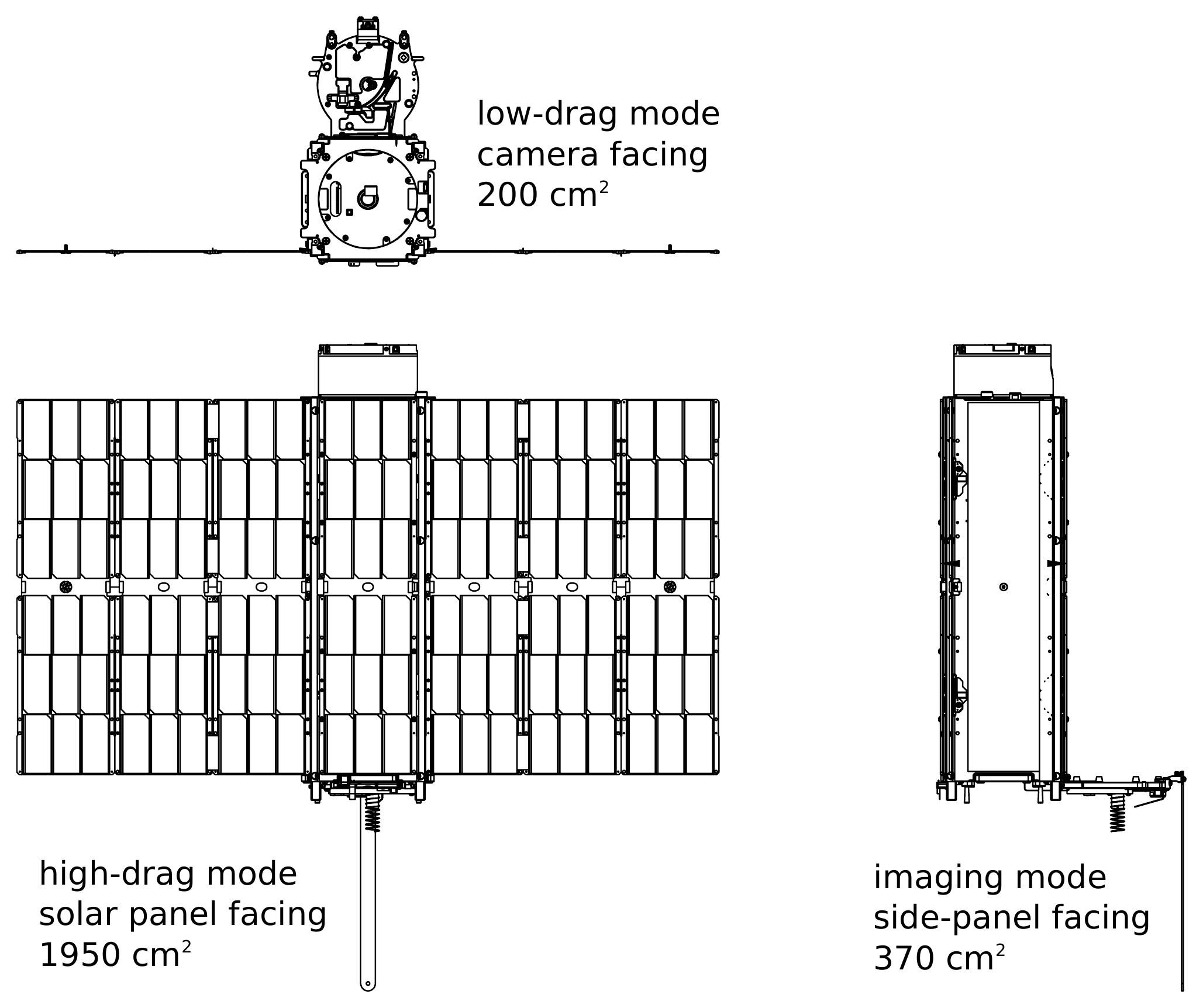} }}%
	\qquad
	\subfloat[High-drag and low-drag attitudes]
	{{\includegraphics[height=2.5in]{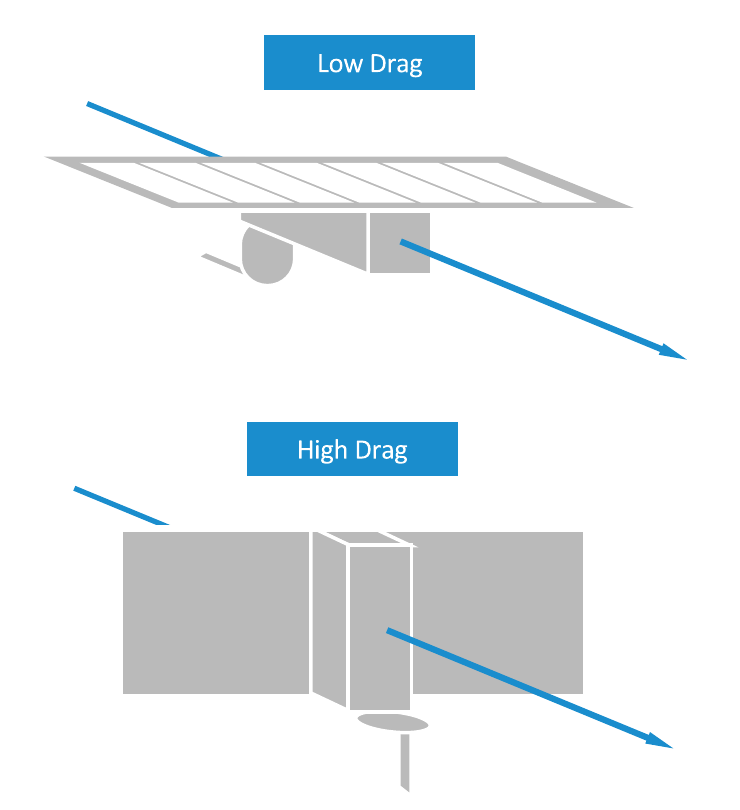} }}%
	\caption{Attitude modes of Dove satellite enable large drag area ratios}
	\label{fig:attitude_modes}%
\end{figure}

Results from the simulation and on-orbit implementation of the controller presented in Section~\ref{section:controller} are presented here. All simulations and results use Planet's Dove satellites. The various attitude modes of Planet's Dove satellites
exhibit a large difference in cross-sectional area
thanks to their deployed solar panels,
as shown in Figure \ref{fig:attitude_modes}. 

Although there is a 10:1 ratio between solar-panel and telescope facing cross-sectional areas, 
only a 3:1 ratio in orbit-derived ballistic coefficients (BC) is observed in reality.
This reduction in control authority is due to several factors: 
\begin{itemize}
	\item Satellite duty-cycle:
	a satellite only executes the desired low-drag or high-drag mode when it is neither imaging
	nor downlinking. 
	In imaging mode, the satellite presents a cross-sectional area between low and high-drag areas, 
	while during downlinks the satellite tracks the ground station with its antenna 
	boresight (co-aligned with telescope).
	\item Attitude pointing accuracy:
	attitude errors effectively reduce the ratio of high to low-drag cross-section areas.
	For example when not imaging or downlinking, the satellites are not using the star camera 
	for attitude determination, resulting in deviations from modeled area.
	\item Skin friction effects:
	even in a perfect low-drag attitude, the large solar panel faces parallel to the velocity
	vector still contribute to drag via skin friction effects \citep{skin_friction}, 
	resulting in the low-drag mode having a higher drag coefficient $c_d$.
\end{itemize}

\subsection{Performance of Controller}
The main contributions of the proposed controller is the slot allocator and solving for the trajectories of all satellites as one highly coupled system. The coupling occurs due to the various flip-flop solutions possible requiring conflicting high-drag windows for the reference satellite.

Although the current controller reliably produces a robust solution, 
there are a few possible for improvements: 
a one-step optimizer and 
second-order optimizations. A source of future work is to solve the two optimization steps in one,
by both assigning slots and generating commands in a single optimization loop.
This would account for inter satellite-pair coupling effects,
but it is currently computationally prohibitive to nest the command generator in the slot allocator.   
Finally, it would be of interest to implement second-order objective functions such as
minimizing clustering during early stages of phasing,
and minimizing station-keeping control effort given a specified control box.

\subsection{In-Space Results}

\begin{figure}[!ht]
	\centering\includegraphics[height=2.5in]{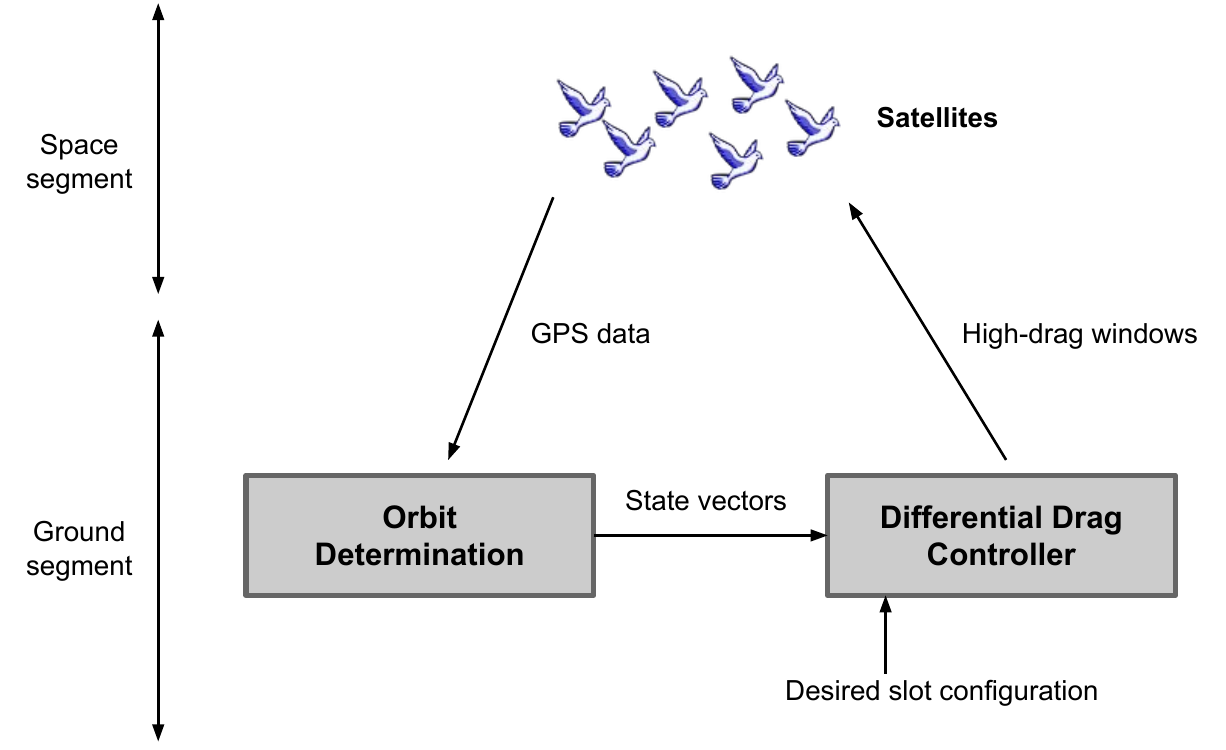}
	\caption[caption]{Orbit determination and differential drag controller runs on the ground}
	\label{fig:system}
\end{figure}

For satellite operations, all estimation and control for differential drag is performed on
the ground (Figure \ref{fig:system}).
GPS data is regularly downloaded during X-band passes and ground-based orbit determination
maintains state vector ephemerides for each satellite.
The differential drag controller then uses these state vectors and the desired
slot configuration to produce a set of high-drag commands that are uploaded to
the satellite.
For Planet's operational orbits in the neighborhood of 500 km altitude,
it is found that a controller update and time discretization of 1-day are sufficient
for achieving desired performance.

\begin{figure}[!ht]
\centering\includegraphics[height=4.0in]{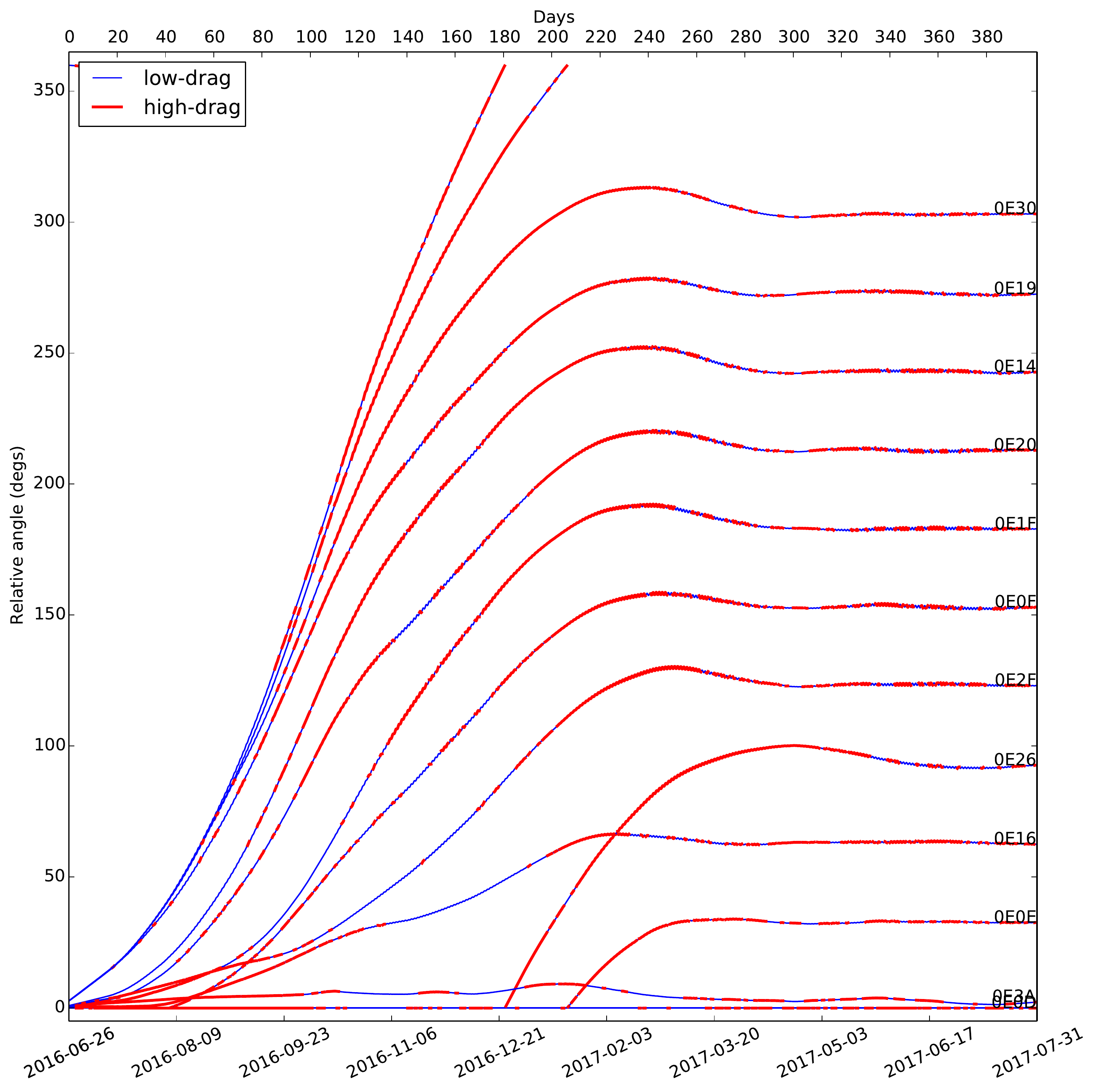}
\caption[caption]{On-orbit results from Flock 2p}
\label{fig:flock2p}
\end{figure}

The differential drag controller described in this paper was applied to Flock 2p,
a set of 12 satellites launched into a circular 510 km sun-synchronous orbit in June 2016.
The satellites were deployed with a non-uniform total spread of 0.5 m/s along-track
$\Delta V$ (deployer ejection speed is 1 m/s, 
along-track spread results from upper stage attitude schedule), 
and the controller allocated high-drag commands
that initially increased this spread before converging to the
desired slot configuration. 
The orbit-derived relative motion and commanded high-drag windows are shown in 
Figure \ref{fig:flock2p}. The same information is presented in a polar plot in Figure~\ref{fig:flock2p_polar} of Appendix A.
This relative motion plot can be recreated using JSpOC TLEs (CATIDs 41606, 41608-41618)
or publicly-available Planet-produced ephemerides\footnote{\url{http://ephemerides.planet-labs.com/}}.
The plan for Flock 2p was to achieve an equally-spaced constellation to
minimize swath overlap and antenna conflicts, except for a pair of satellites
that would maintain $3.6^{\circ}$ spacing to demonstrate the line scanner imaging strategy.

\begin{figure}%
\centering
\subfloat[Ballistic Coefficients (BC)]{{
\includegraphics[width=.45\linewidth]{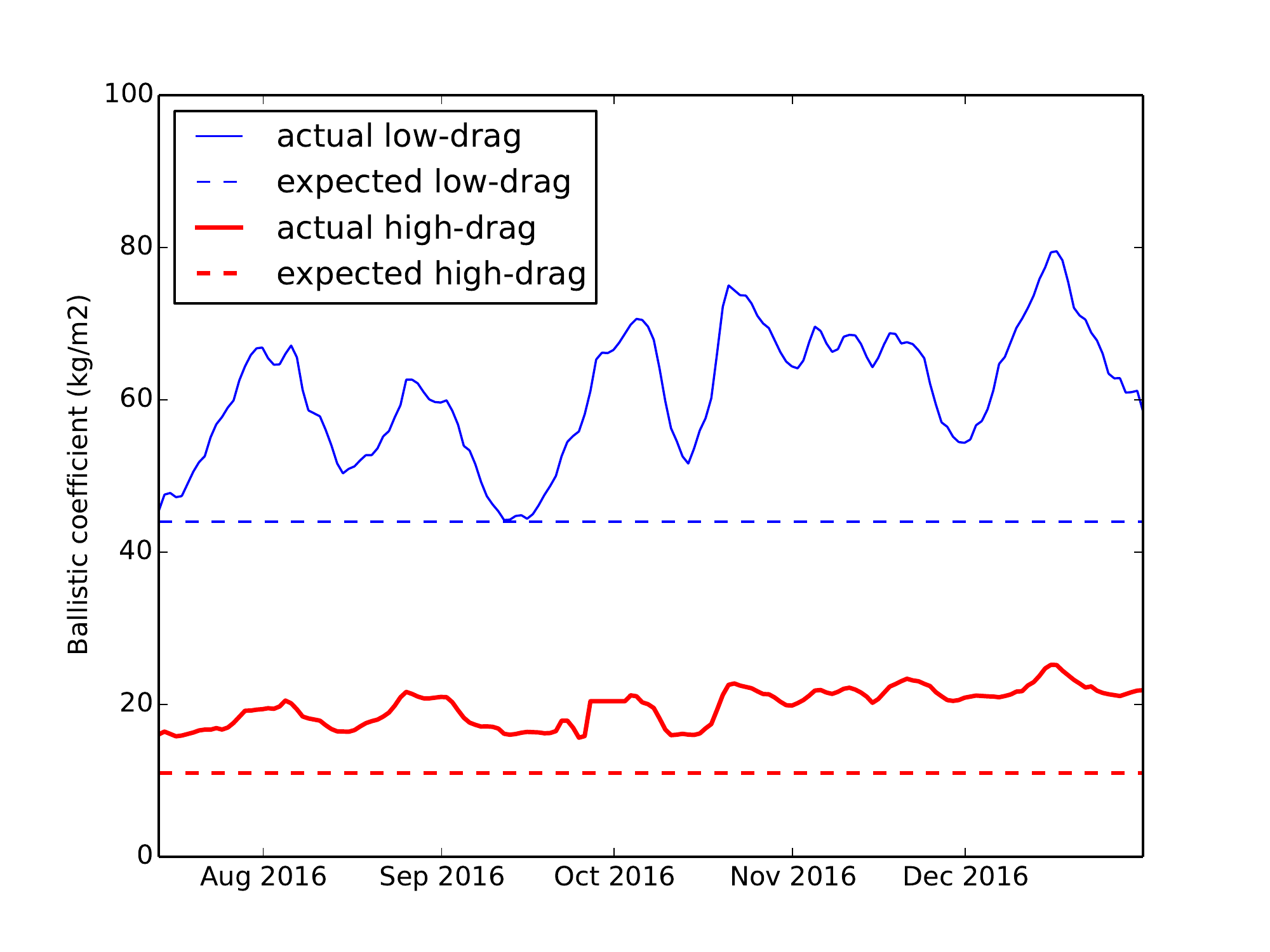} }}%
\qquad
\subfloat[Control authority]{{
\includegraphics[width=.45\linewidth]{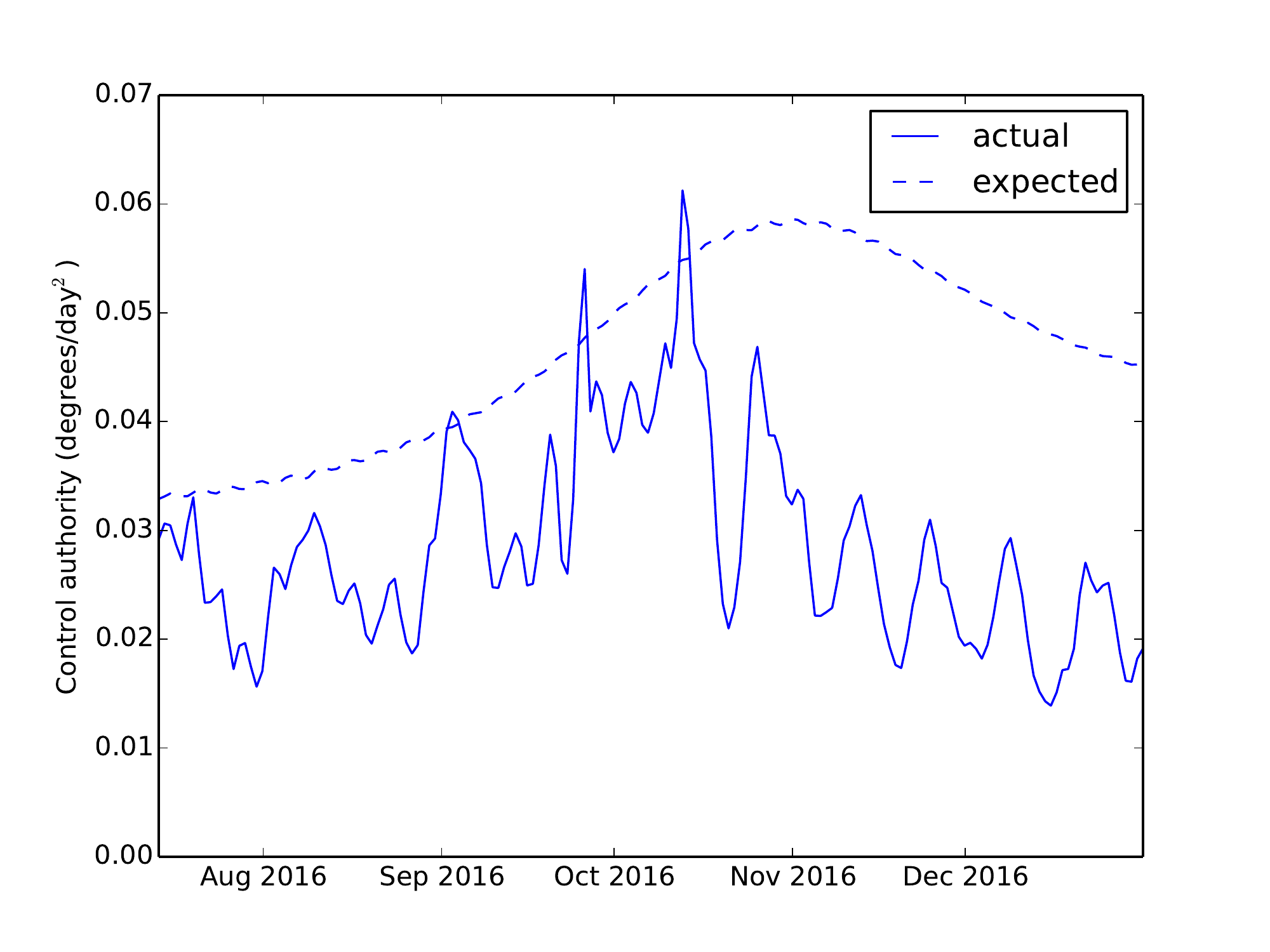} }}%
\caption{Expected vs. actual atmosphere-dependend performance}%
\label{fig:BC_vs_t}%
\end{figure}

The primary lesson learned from automated on-orbit control of Flock 2p is to account
for the significant unmodeled variations in effective BC.
These fluctuations have been treated extensively by others \citep{vallado_atm}, 
and its specifics as encountered by Planet's differential drag phasing are discussed below.
Figure \ref{fig:BC_vs_t}a shows the expected versus actual BCs;
the expected values were pre-launch estimates based on an older Flock launched into
a 600 km orbit in 2014.
Figure \ref{fig:BC_vs_t}b shows the resulting available control authority 
with pre-launch BCs using the solar flux prediction data at the time, 
and the actual control authority available from post-processed orbit data.
One can see that the actual control authority available was significantly less than 
that predicted, by up to 50\% 6 months after launch. 
This underestimation of control authority is responsible for the overshoot observed
in Figure \ref{fig:flock2p};
and caused some churn with the slot allocator as
it attempted to continuously target slots using its optimistic estimate of control authority.

Even though the satellites' attitude control systems were consistent 
at maintaining low and high-drag modes,
the resulting BCs exhibit large periodic fluctuations.
The same fluctuating behavior has also been observed across year-long time-scales
on satellites that are tumbling at rates much shorter than orbit fit-spans 
(therefore effectively presenting a constant cross-sectional area).
One would expect the BCs in various consistent modes to be time-invariant
when using a-posteriori solar flux data and industry standard atmosphere models 
like MSIS00 or Jacchia-Bowman 2008 \citep{jb2008},
so the fluctuating behavior of orbit-derived BCs is attributed to unmodeled
atmospheric density variations due to solar phenomena
(the period of the fluctuations also matches the Sun's synodic period of 26 days).
Atmosphere models have evolved in complexity, but they still fundamentally rely on 
a handful or less of channels of solar data.
Since the Sun is the greatest contributor to these atmospheric density
fluctuations, there must be potential for improved atmosphere
models that make use of more data from the Sun, 
perhaps 10's or 100's channels and at higher frequencies
from various measurement platforms that were not historically available.

The mitigation was to continuously update the controller's estimate of 
low and high-drag BCs based on observed control authority from satellites in those modes.

\FloatBarrier

\subsection{Future Contellations}

\begin{figure}[!ht]
\centering\includegraphics[height=4.0in]{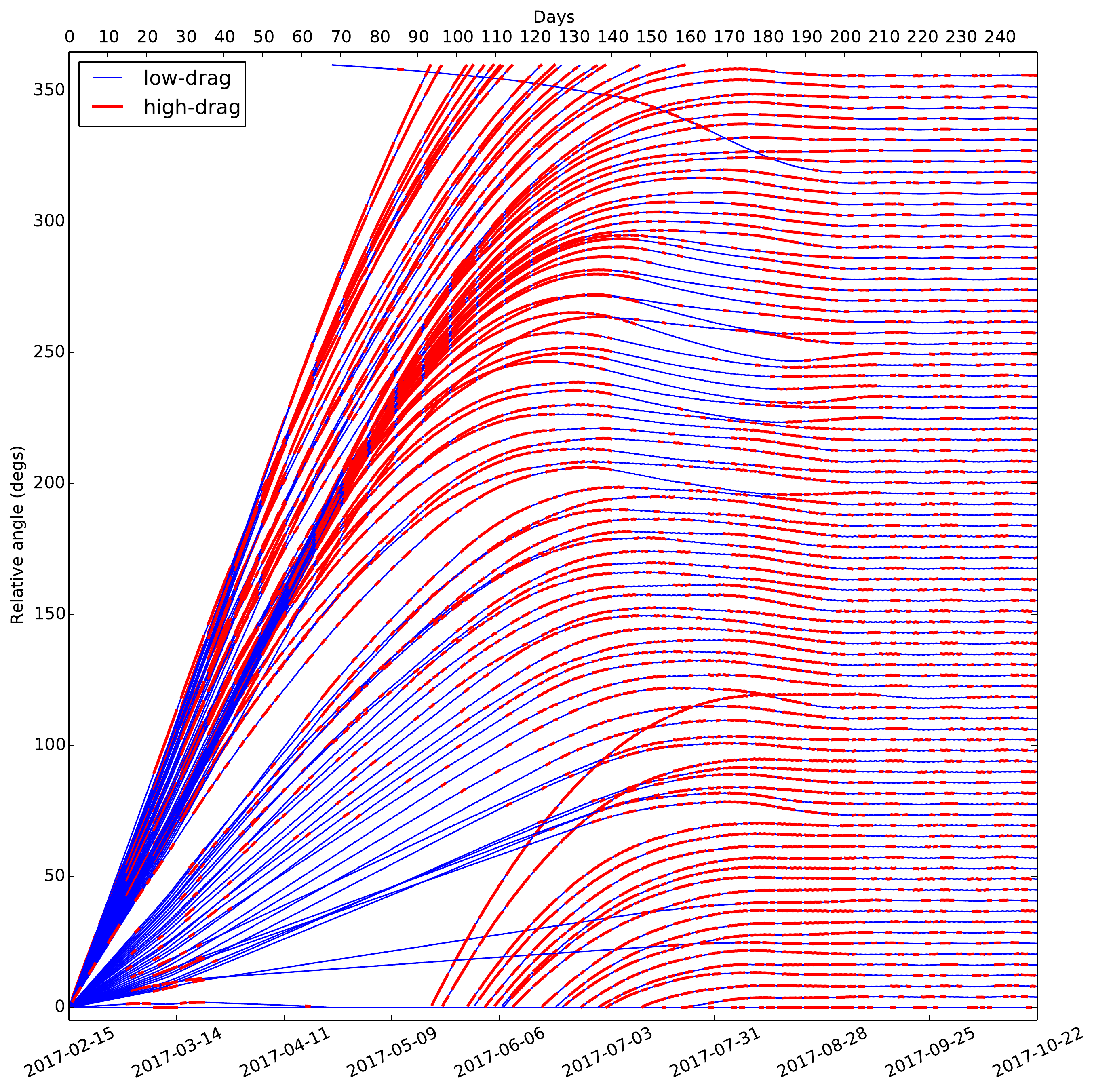}
\caption[caption]{Simulation of Flock 3p phasing and station-keeping}
\label{fig:flock3p}
\end{figure}

Planet recently launched Flock 3p on February 14th 2017 into a 505 km altitude orbit.
It consists of 88 Doves that were individually deployed and targeted to
achieve a roughly uniform distribution of along-track 
$\Delta V$'s. 
The total initial spread is 2 m/s, spanning two groups of satellites,
with a gap in the middle consisting of 13 non-Planet satellites.
Figure \ref{fig:flock3p} shows the simulated behavior of Flock 3p from
deployment through station-keeping. The same information is presented in a polar plot in Figure~\ref{fig:flock3p_polar} of Appendix A.
The simulation factors in the fact that not all satellites are eligible to perform 
differential drag maneuvers right-away,
as commissioning of specific satellites are staggered over the first 80 days.

\FloatBarrier

\section{Conclusion}\label{section:conclusion}

This paper details a controller design for phasing and station-keeping large
fleets of satellites with only differential drag.
The controller works in three steps: 
relative motion estimation, slot allocating and high-drag command generation for the entire couple system.

Relative motion is first estimated in the along-track direction to obtain
the mean angular separation angle and speed between each satellite
and a designated reference. 
The satellites are assumed to be commandable into discrete low and high-drag modes,
and angular acceleration for each mode is evaluated across a time horizon of interest
to capture variations in mean atmospheric density.

Two optimization problems are then solved sequentially: 
the first to assign each satellite to a specific constellation slot and 
the second to generate time-discretized commands to guide each satellite
to its slot using a coupled system.
Both optimization problems seek to minimize the required phasing time
given the available control authority, 
and are solved with simulated annealing,
a method that lends well to the discretized nature of the problem.
They also make use of the solution to the two-satellite sub-problem to
either estimate the required phasing time or provide an initial guess to 
the general multi-satellite problem. 
Although both components are required for an optimal solution, the coupling of the system
has a larger influence on the optimality and feasibility of the solutions. 

The closed-loop performance of the controller is demonstrated 
with an operational fleet of twelve satellites, 
and it is also being applied to another fleet of 88 satellites that was
recently launched in February 2017.

\section{Acknowledgments}
The authors would like to acknowledge Planet's Missions and Spacecraft Design teams
for putting together all the pieces needed to execute the automated differential drag
operations described in the paper.

\appendix
\section*{Appendix}
\subsection{Polar Plots}

\begin{figure}[!ht]
\centering\includegraphics[height=3.5in]{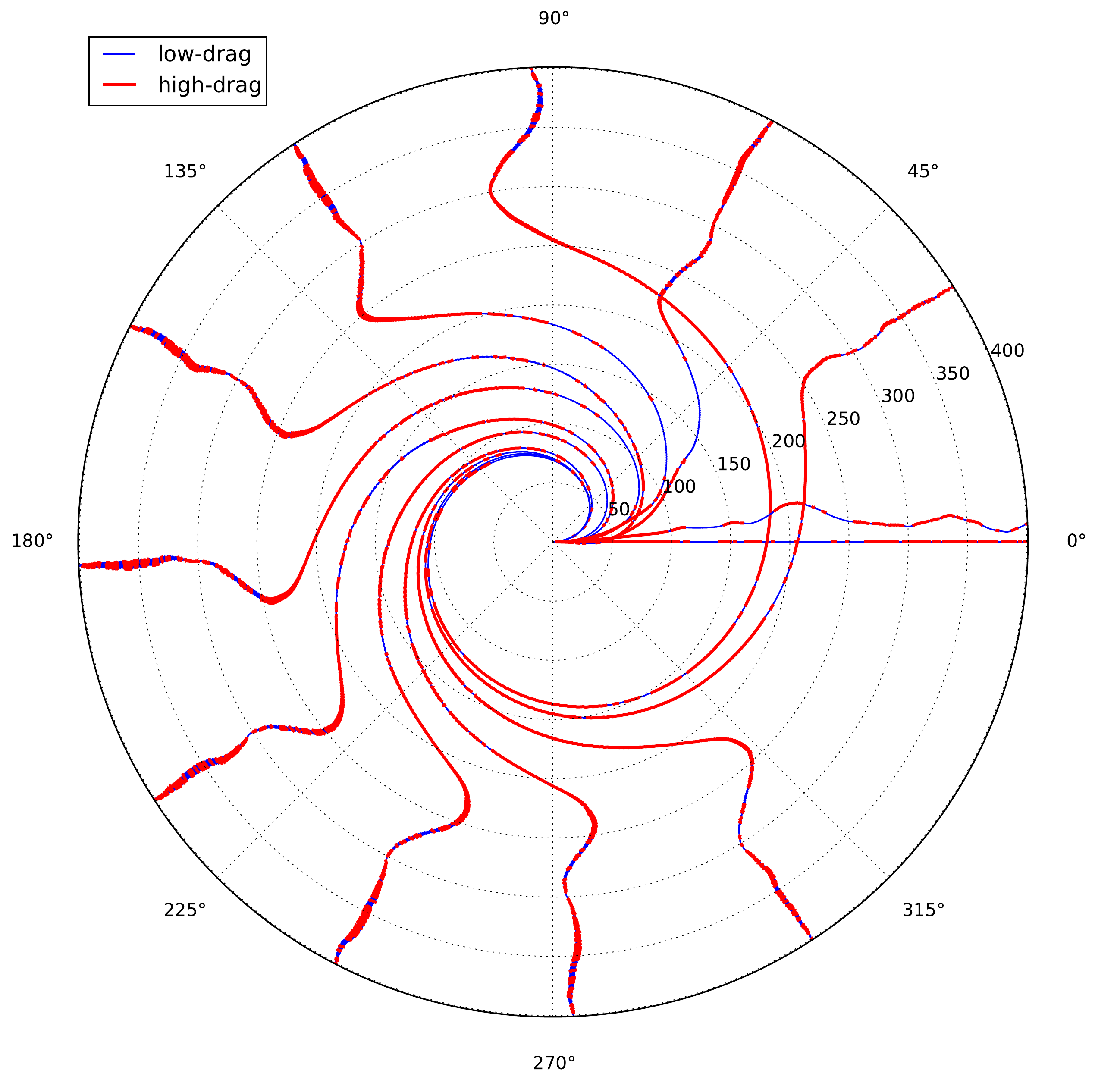}
\caption[caption]{On-orbit results from Flock 2p}
\label{fig:flock2p_polar}
\end{figure}

\begin{figure}[!ht]
\centering\includegraphics[height=3.5in]{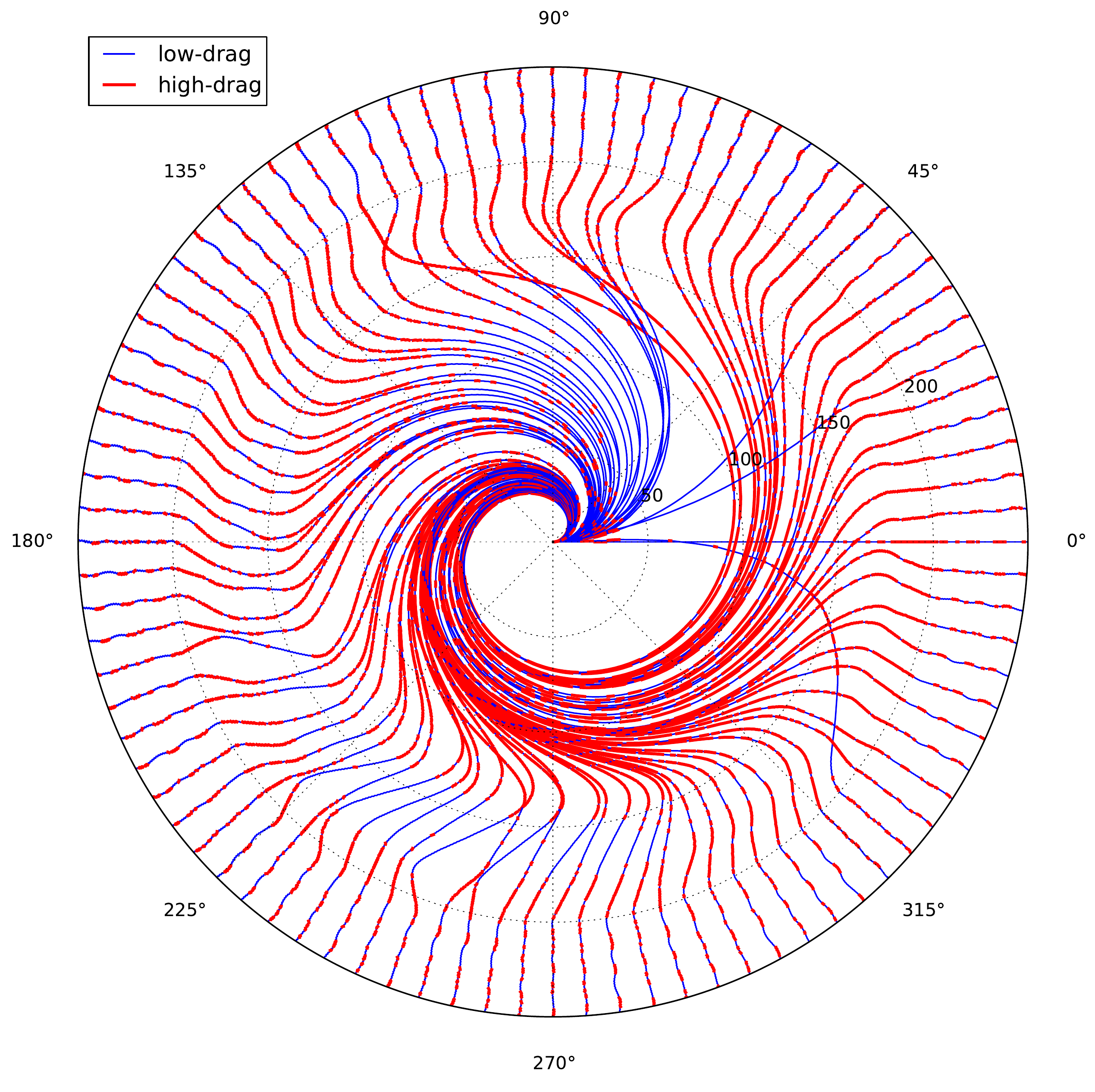}
\caption[caption]{Simulation of Flock 3p phasing and station-keeping}
\label{fig:flock3p_polar}
\end{figure}

Figures \ref{fig:flock2p_polar} and \ref{fig:flock3p_polar}
are polar projections of Figures \ref{fig:flock2p} and \ref{fig:flock3p} respectively.
Radial axis shows time (in days), while relative angle is on the $\theta$ axis.
These polar projections better illustrate the relative angles that wrap around,
at the cost of a time axis that is harder to read.

\FloatBarrier
    
\subsection{Analytic solution for two satellite case}

This section presents the analytical solution to the time-optimal 
open-loop differential drag control of two satellites 
under time-invariant control authority $\ddot{\theta}$.
The initial relative motion is given by
$\begin{bmatrix} \theta & \dot{\theta} \end{bmatrix}^T_0$
and the final condition by
$\begin{bmatrix} \theta & \dot{\theta} \end{bmatrix}^T_f$.

The solution consists of two phases 
(Figure \ref{fig:flipflop}):
one satellite high-drags for a duration $\Delta t_A$ while the other low-drags,
then the roles reverse for another duration $\Delta t_B$.
If the reference high-drags first, 
the effective control authority during phases A and B are 
$\ddot{\theta}_A = -\ddot{\theta}$ and 
$\ddot{\theta}_B = \ddot{\theta}$ respectively.
Similarly if the satellite high-drags first, then one has
$\ddot{\theta}_A = \ddot{\theta}$ and 
$\ddot{\theta}_B = -\ddot{\theta}$ respectively.

The relative motion at the mid-point between phases A and B is expressed as
$\begin{bmatrix} \theta & \dot{\theta} \end{bmatrix}^T_m$:

\begin{equation}
\label{eq:flipflop_analytic_A}
\begin{split}
\theta_m &= \theta_0 + \dot{\theta}_0 \Delta t_A + \frac{1}{2} \ddot{\theta}_A \Delta t_A^2 \\
\dot{\theta}_m &= \dot{\theta}_0 + \ddot{\theta}_A \Delta t_A
\end{split}
\end{equation}

The final condition after phase B is then expressed as:

\begin{equation}
\label{eq:flipflop_analytic_B}
\begin{split}
\theta_f &= \theta_m + \dot{\theta}_m \Delta t_B + \frac{1}{2} \ddot{\theta}_B \Delta t_B^2 \\
\dot{\theta}_f &= \dot{\theta}_m + \ddot{\theta}_B \Delta t_B
\end{split}
\end{equation}

$\Delta t_A$ and $\Delta t_B$ are solved for from equations \ref{eq:flipflop_analytic_A}
and \ref{eq:flipflop_analytic_B} by first eliminating $\theta_m$ and $\dot{\theta}_m$,
and introducing the short hands
$\Delta \theta = \theta_f - \theta_0$
and
$\Delta \dot{\theta} = \dot{\theta}_f - \dot{\theta}_0$,
to reveal $\Delta t_B$ as a function of $\Delta t_A$:

\begin{equation}
\label{eq:flipflop_analytic_dtB}
\Delta t_B = \frac{\Delta \dot{\theta} - \ddot{\theta}_A \Delta t_A}{\ddot{\theta}_B}
\end{equation}

and $\Delta t_A$ from the quadratic:

\begin{equation}
\label{eq:flipflop_analytic_dtA}
\Bigg[ \frac{1}{2} \ddot{\theta}_A \bigg( \frac{\ddot{\theta}_A}{\ddot{\theta}_B} - 1 \bigg) \Bigg] \Delta t_A^2
+ \Bigg[ \dot{\theta}_0 \bigg( \frac{\ddot{\theta}_A}{\ddot{\theta}_B} - 1 \bigg) \Bigg] \Delta t_A
+ \Bigg[ \Delta \theta - \frac{\Delta \dot{\theta}}{\ddot{\theta}_B} 
\bigg( \dot{\theta}_0 + \frac{1}{2} \Delta \dot{\theta} \bigg) \Bigg] = 0 \\
\end{equation}
    
There are four possible solutions to the ($\Delta t_A$, $\Delta t_B$) couple
given the ambiguity of which satellite should high-drag first, 
and the two roots to Equation \ref{eq:flipflop_analytic_dtA}, 
but only one couple has positive real values, the physical solution to the problem.

\bibliographystyle{AAS_publication}   
\bibliography{paper_diffdrag}

\end{document}